\newbox\tempa
\newbox\tempb
\newdimen\tempc
\newbox\tempd

\def\mud#1{\hfil $\displaystyle{#1}$\hfil}
\def\rig#1{\hfil $\displaystyle{#1}$}

\def\inruleanhelp#1#2#3{\setbox\tempa=\hbox{$\displaystyle{\mathstrut #2}$}%
                        \setbox\tempd=\hbox{$\; #3$}%
                        \setbox\tempb=\vbox{\halign{##\cr
        \mud{#1}\cr
        \noalign{\vskip\the\lineskip}%
        \noalign{\hrule height 0pt}%
        \rig{\vbox to 0pt{\vss\hbox to 0pt{\copy\tempd \hss}\vss}}\cr
        \noalign{\hrule}%
        \noalign{\vskip\the\lineskip}%
        \mud{\copy\tempa}\cr}}%
                      \tempc=\wd\tempb
                      \advance\tempc by \wd\tempa
                      \divide\tempc by 2 }

\def\inrulean#1#2#3{{\inruleanhelp{#1}{#2}{#3}%
                     \hbox to \wd\tempa{\hss \box\tempb \hss}}}

\def\lowerhalf#1{\hbox{\raise -0.8\baselineskip\hbox{#1}}}

\def\bnfas{\mathrel{::=}}

\def\lam{\lambda}

\def\ldot{\mathord{.}\;}
\long\def\ignore#1{}

\renewcommand{\cite}{\citep}

\newcommand{\crel}[2]{#1\sim#2}

\newcommand{\termeq}[2]{\mathsf{aeq}\;#1\;#2}

\newcommand{\istm}[1]{\mathsf{is\_tm}\;#1}

\newcommand{\termequal}[2]{\mathsf{deq}\;#1\;#2}
\newcommand{\termeqcon}{\mathsf{aeq}}
\newcommand{\termequalcon}{\mathsf{deq}}
\newcommand{\app}[2]{\mathsf{app}\; #1\; #2}

\renewcommand{\lam}[2]{\mathsf{lam}\,#1.\,#2}

\newcommand{\lamt}[3]{\mathsf{lam}\,#1^{}.\,#3}

\newcommand{\alt}{\, \ensuremath{\mathsf{+}} \, }

\newcommand{\ictx}{\Phi_x}
\newcommand{\isch}{S_x}
\newcommand{\actx}{\Phi_{xa}}
\newcommand{\asch}{S_{xa}}
\newcommand{\dctx}{\Phi_{xd}} 
\newcommand{\dsch}{S_{xd}}

\documentclass[submission,copyright,creativecommons]{eptcs}
\providecommand{\event}{LFMTP 2015} 
\usepackage{breakurl}             

\usepackage{proof}
\usepackage{cite}

\newtheorem{theorem}{Theorem}[section]

\title{An Open Challenge Problem Repository for Systems Supporting Binders}
\author{Amy Felty 
\institute{School of Electrical Engineering and\\ Computer
  Science\\University of Ottawa\\Ottawa, Canada} 
 \email{afelty@eecs.uottawa.ca}
  \and
  Alberto Momigliano
  \institute{Dipartimento di Informatica\\Universit\`{a} degli Studi
    di Milano\\Milano, Italy}
  \email{momigliano@di.unimi.it}
  \and
  Brigitte Pientka
  \institute{School of Computer Science\\McGill University\\Montreal, Canada}
  \email{bpientka@cs.mcgill.ca}
}

\def\titlerunning{ORBI}
\def\authorrunning{A. Felty, A. Momigliano \& B. Pientka}
\begin{document}
\maketitle

\begin{abstract}
  A variety of logical frameworks support the use of higher-order
  abstract syntax in representing formal systems; however, each system
  has its own set of benchmarks. Even worse, general proof assistants
  that provide special libraries for dealing with binders offer a
  very limited evaluation of such libraries, and the examples given
  often do not exercise and stress-test key aspects that arise in the
  presence of binders.  In this paper we design an open repository
  \emph{ORBI} (\underline{O}pen challenge problem
  \underline{R}epository for systems supporting reasoning with
  \underline{BI}nders).  We believe the field of reasoning about
  languages with binders has matured, and a common set of benchmarks
  provides an important basis for evaluation and qualitative
  comparison of different systems and libraries that support binders,
  and it will help to advance the field.
\end{abstract}

\section{Introduction}

A variety of logical frameworks support the use of higher-order
abstract syntax (HOAS) in representing formal systems; however, each
system has its own set of benchmarks, often encoding the same object
logics with minor differences. Even worse, general proof assistants
that provide special libraries for dealing with binders often offer
only a very limited evaluation of such libraries, and the examples
given often do not exercise and stress-test key aspects that arise in
the presence of binders.

The \textsc{PoplMark} challenge~\cite{Aydemir05TPHOLS} was an
important milestone in surveying the state of the art in mechanizing
the meta-theory of programming languages. We ourselves proposed
several specific benchmarks~\cite{companion} that are \emph{crafted}
to highlight the differences between the designs of various
meta-languages with respect to reasoning with and within a context of
assumptions, and we compared their implementation in four systems: the
logical framework Twelf~\cite{TwelfSP}, the dependently-typed
functional language Beluga~\cite{Pientka:IJCAR10,Pientka:CADE15}, the
two-level Hybrid system~\cite{MMF07,FeltyMomigliano:JAR10} as
implemented on top of Coq and Isabelle/HOL, and the Abella
system~\cite{Gacek:IJCAR08}.  Finally, several systems that support
reasoning with binders, in particular systems concentrating on
modeling binders using HOAS, also provide a large collection of
examples and case studies. For example, Twelf's wiki
(\url{http://twelf.org/wiki/Case_studies}), Abella's library
(\url{http://abella-prover.org/examples}), Beluga's distribution, and
the Coq implementation of Hybrid
(\url{http://www.site.uottawa.ca/~afelty/HybridCoq/}) contain sets of
examples that highlight the many issues surrounding binders.

As the field matures, we believe it is important to be able to systematically and
qualitatively evaluate approaches that support reasoning with binders. Having
benchmarks is a first step in this direction. In this paper, we propose a
common infrastructure for representing challenge problems and a central,
\underline{O}pen challenge problem \underline{R}epository for systems supporting
reasoning with \underline{BI}nders ({ORBI}) for sharing benchmark
problems based on the notation we have developed. 

ORBI is designed to be a human-readable, easily machine-parsable,
uniform, yet flexible and extensible language for writing
specifications of formal systems including grammar, inference rules,
contexts and theorems. The language directly upholds HOAS
representations and is oriented to support the mechanization of 
benchmark problems in Twelf, Beluga, Abella, and Hybrid, without
hopefully precluding other existing or future HOAS systems. At the
same time, we hope it also is amenable to translations to systems
using other representation techniques such as nominal ones.

We structure the language in two parts:
 \begin{enumerate}
 \item the problem description, which includes the grammar of
   the object language syntax, inference rules, context schemas, and
   context relations
 \item the logic language, which includes syntax for expressing
   theorems and directives to ORBI2X\footnote{Following TPTP's
     nomenclature \cite{TPTP}, we call ``ORBI2X'' any tool taking an
     ORBI specification as input; for example, the translator for
     Hybrid described in \cite{HabliFelty:PXTP13} turns syntax, inference rules, and
     context definitions of ORBI into input to the Coq version of
     Hybrid, and it is designed so that it can be adapted fairly directly
     to output Abella scripts.}  tools.
 \end{enumerate}

 We begin in Sect.~\ref{sec:example} with a running example.  We
 consider the untyped lambda-calculus as an object logic (OL), and
 present the syntax, some judgments, and sample theorems.  In
 Sect.~\ref{sec:orbi}, we present ORBI by giving its grammar and
 explaining how it is used to encode our running example;
 Sect.~\ref{ssec:spec} and Sect.~\ref{ssec:logic} present the two
 parts of this specification as discussed above.
 We discuss related work in Sect.~\ref{sec:related}.

 We consider the notation that we present here as a first attempt at
 defining ORBI (Version 0.1), where the goal is to cover the
 benchmarks considered in~\cite{companion}.  As new benchmarks are
 added, we are well aware that we will need to improve the syntax and
 increase the expressive power---we discuss limitations and some
 possible extensions in Sect.~\ref{sec:concl}.

\section{A Running Example}
\label{sec:example}

The first question that we face when defining an OL is how to describe
well-formed objects.  Consider the untyped lambda-calculus, defined by
the following grammar:
$$M \bnfas  x \mid \lamt{x}{A}{M} \mid \app{M_1}{M_2}.$$

To capture additional information that is often useful in proofs, such
as when a given term is \emph{closed}, it is customary to give
inference rules in natural deduction style for well-formed terms using
hypothetical and parametric judgments.
However, it is often convenient to present hypothetical judgments in a
\emph{localized} form, reducing some of the ambiguity of the
two-dimensional natural deduction notation, and providing more
structure. We therefore introduce an \emph{explicit} context for
bookkeeping, since when establishing properties about a given system,
it allows us to consider the variable case(s) separately and to state
clearly when considering closed objects, i.e., an object in the empty
context. More importantly, while structural properties of contexts are
implicitly present in the natural deduction presentation of inference
rules (where assumptions are managed informally), the explicit context
presentation makes them more apparent and highlights their use in
reasoning about contexts.
\[
\begin{array}{l}
\infer[tm_{v}]{\Gamma \vdash \istm{x}}{\istm{x} \in \Gamma}
\quad
\infer[tm_{l}]{\Gamma \vdash \istm{(\lamt{x}{A}{M}})}{
& 
\Gamma, \istm{x} \vdash \istm{M}}
\qquad
 \infer[tm_a]{\Gamma \vdash \istm{(\app{M_1}{M_2})}}
 {\Gamma \vdash \istm{M_1}\qquad  \Gamma \vdash \istm{M_2}}
\\[1em]
 \end{array}
\]

Traditionally, a context of assumptions is characterized as a sequence
of formulas $A_1, A_2, \ldots, A_n$ listing its elements separated by
commas~\cite{PierceFirstBook,GirardLafontTaylor:proofsAndTypes}.
In~\cite{orbi}, we argue that this is not expressive enough to capture
the structure present in contexts, especially when mechanizing object
logics, and we define context \emph{schemas} to introduce the required
extra structure:
\[
\begin{array}{lrcl}
&\mbox{Atom} \quad A & & \\
&\mbox{Block of declarations} \quad D & \bnfas & A  \mid D; A\\
&\mbox{Context} \quad \Gamma & \bnfas & \cdot \mid \Gamma , D\\
&\mbox{Schema} \quad S & \bnfas & D_s \mid D_s \alt S
\end{array}
\]
A context is a sequence of declarations $D$ where a declaration is a
block of individual atomic assumptions separated by
'$;$'.\footnote{This is an oversimplification, since there are
  well-known specifications where contexts have more structure, see  the
  solution to the \textsc{PoplMark} challenge in \cite{Pientka07} and
the examples in \cite{hoabella}. In fact, those are already legal ORBI specs.}
The '$;$' binds tighter than '$,$'. We treat contexts as ordered,
i.e., later assumptions in the context may depend on earlier ones, but
not vice versa---this in contrast to viewing contexts as multi-sets.
Just as types classify terms, a \emph{schema} will classify meaningful
structured sequences. A schema consists of declarations $D_s$, where
we use the subscript $s$ to indicate that the declaration occurring in
a concrete context having schema $S$ may be an \emph{instance} of
$D_s$. We use $\alt$ to denote the alternatives in a context schema.
For well-formed terms, contexts have a simple structure where each
block contains a single atom, expressed as the following schema
declaration:
$$\isch := \istm{x}.$$
We write $\ictx$ to represent a context \emph{satisfying} schema $\isch$ (and
similarly for other context schemas appearing in this paper).
Informally, this means that $\ictx$ has the form
$\istm{x_1},\ldots,\istm{x_n}$ where $n\ge0$ and $x_1,\ldots,x_n$ are
distinct variables.  (See~\cite{orbi} for a more formal account.)

For our running example, we consider two more simple judgments.
The first is \emph{algorithmic equality} for the untyped
lambda-calculus, written $(\termeq{M}{N})$.  We say that two terms are
algorithmically equal provided they have the same structure with
respect to the constructors.
$$
\infer[ae_v]{\Gamma \vdash \termeq{x}{x}}{\termeq{x}{x} \in \Gamma}
\qquad 
\infer[ae_l]{\Gamma \vdash \termeq{(\lam{x}{M})}{(\lam{x}{N})}}{
\Gamma, \istm{x}; \termeq{x}{x} \vdash \termeq{M}{N}} 
\qquad
\infer[ae_a]{\Gamma \vdash \termeq{(\app{M_1}{M_2})}{(\app{N_1}{N_2})}}
{\Gamma \vdash \termeq{M_1}{N_1} & \Gamma \vdash \termeq{M_2}{N_2}}
$$
The second is \emph{declarative equality} written
$(\termequal{M}{N})$, which includes versions of the above three rules
called $de_v$, $de_l$, and $de_a$, where $\termeqcon$ is replaced by
$\termequalcon$ everywhere, plus reflexivity, symmetry and
transitivity shown below.\footnote{We acknowledge that this definition
  of declarative equality has a degree of redundancy: the assumption
  $\termequal x x$ in rule $de_l$ is not needed, since rule $de_r$
  plays the variable role. However, this formulation exhibits issues,
  such as \emph{context subsumption},
  that would otherwise require more complex benchmarks.}
$$
\infer[de_r]{\Gamma \vdash \termequal{M}{M}}{}
\quad
\infer[de_s]{\Gamma \vdash \termequal{M}{N}}{
\Gamma \vdash \termequal{N}{M}}
\quad
\infer[de_t]{\Gamma \vdash \termequal{M}{N}}{
\Gamma \vdash \termequal{M}{L} & \Gamma \vdash \termequal{L}{N}}
$$
These judgments give rise to the following schema declarations:
$$\begin{array}[t]{lll}
\asch & := & \istm{x}; \termeq{x}{x} \\
\dsch & := & \istm x;\termequal{x}{x}\\
S_{da} & := & \istm x; \termequal{x}{x} ;\termeq x x
\end{array}$$
The first two come directly from the $ae_l$ and $de_l$ rules where
declaration blocks come in pairs.  The third combines the two, and is
used below in stating one of the example theorems.

When stating properties, we often need to relate two judgments 
to each other, where each one has its own context.  For
example, we may want to prove statements such as ``if $\ictx \vdash
J_1$ then $\actx \vdash J_2$.'' The proofs in~\cite{companion} use two
approaches.\footnote{In proofs on paper, the differences between the
  two approaches usually do not appear; they are present in the details
  that are left implicit, but must be made explicit when mechanizing
  proofs. For example, on-paper versions of the admissibility of
  reflexivity that make these distinctions explicit appear
  in~\cite{orbi} as proofs of Theorems 7 and 8.}
In the first, the statement is reinterpreted in the
\emph{smallest} context that collects all relevant assumptions;  we
call this the \emph{generalized context} approach (G). The above
statement becomes ``if $\actx \vdash
J_1$ then $\actx \vdash J_2$.''  
As an example theorem, we consider the completeness of declarative 
equality with respect to  algorithmic equality, of which we only show the
interesting left-to-right direction.
\begin{theorem}[Completeness, G Version ]$\;$
  \begin{description}
  \item[Admissibility of Reflexivity]
         If $\actx \vdash \istm M$ then $\actx \vdash \termeq M M$.
  \item[Admissibility of Symmetry]
         If $\actx \vdash \termeq M N$ then
         $\actx \vdash \termeq N M$.
\item[Admissibility of Transitivity]
         If $\actx \vdash \termeq M N$ and $\actx \vdash \termeq N L$ then
         $\actx \vdash \termeq M L$.
  \item[Main Theorem] If $\Phi_{da} \vdash \termequal{M}{N}$ then $\Phi_{da} \vdash \termeq M N$.
  \end{description}
\end{theorem}

In the second approach, we
state how two (or more) contexts are related via context relations. For example, the following
relation captures the fact that $\istm x$ will occur in $\ictx$ in sync with an
assumption block containing $\istm{x}; \termeq x x$ in $\actx$.
\[
\begin{array}{c}
\infer[]{\crel \ldot \ldot}{} \quad\quad\quad\quad
\infer[]{\crel {\ictx, \istm x} {\actx, \istm{x}; \termeq{x}{x}}}{\crel \ictx \actx}
\end{array}
\]
Similarly, we can define $\crel \actx \dctx$.
\[
\begin{array}{c}
\infer[]{\crel \ldot \ldot}{} \quad\quad\quad\quad
\infer[]{\crel {\actx, \istm{x};\termeq x x}
                 {\dctx,  \istm x; \termequal x x}}{\crel \actx \dctx} 
\end{array}
\]
We call this the \emph{context relations} approach (R).  The theorems are then typically stated
as: ``if $\ictx \vdash J_1$ and $\crel \ictx  \actx$ then $\actx \vdash
J_2$.'' We can then revisit the completeness theorem for algorithmic equality
together with the necessary lemmas as follows.

\begin{theorem}[Completeness, R Version]$\;$
  \begin{description}
  \item[Admissibility of Reflexivity] 
Assume $\crel \ictx \actx$.
         If $\ictx \vdash \istm M$ then $\actx \vdash \termeq M M$.
  \item[Admissibility of Symmetry]
         If $\actx \vdash \termeq M N$ then
         $\actx \vdash \termeq N M$.

  \item[Admissibility of Transitivity]
         If $\actx \vdash \termeq M N$ and $\actx \vdash \termeq N L$ then
         $\actx \vdash \termeq M L$.
  \item[Main Theorem]
  Assume $\crel \actx \dctx$.
 If $\Phi_{xd} \vdash \termequal{M}{N}$ then $\Phi_{xa} \vdash \termeq M N$.
  \end{description}
\end{theorem}

\section{ORBI}
\label{sec:orbi}
ORBI aims to provide a common framework for systems that support
reasoning with binders.
Currently, our design is geared towards systems supporting HOAS,
where there are (currently) two main
approaches. On one side of the spectrum  we have systems that
implement various dependently-typed calculi.
Such systems include Twelf, Beluga, and
Delphin~\cite{Schuermann:LFMTP08}. All these systems also provide, to
various degrees, built-in support for reasoning modulo structural
properties of a context of assumptions. These systems support inductive
reasoning over terms as well as rules. Often it is more elegant in these
systems to state theorems using the G-version~\cite{companion}.

On the other side there are systems based on a proof-theoretic
foundation, which typically follow a two-level approach: they implement a
specification logic (SL) inside a higher-order logic or type
theory. Hypothetical judgments of object languages are modeled using
implication in the SL and parametric judgments are handled via
(generic) universal quantification.  Contexts are commonly represented
explicitly as lists or sets in the SL, and structural properties are
established separately as lemmas. For example substituting for an
assumption is justified by appealing to the cut-admissibility lemma of
the SL\@.  These lemmas are not directly and intrinsically supported
through the SL, but may be integrated into a system's automated
proving procedures, usually via tactics. Induction is usually only
supported on derivations, but not on terms. Systems following this
philosophy 
include Hybrid and Abella.
Often these systems are better suited to proving R-versions of theorems.

The desire for ORBI to cater to both type and proof theoretic
frameworks requires an almost impossible balancing act between the two
views.  For example, contexts are first-class and part of the
specification language in Beluga; in Twelf, schemas for contexts are
part of the specification language, which is an extension of LF, but
users cannot explicitly quantify over contexts and manipulate them as
first-class objects; in Abella and Hybrid, contexts are (pre)defined
using inductive definitions on the reasoning level. We will describe
next our common infrastructure design, directives, and guidelines
that allow us to cater to
existing systems supporting HOAS.

\subsection{Problem Description in ORBI}
\label{ssec:spec}

ORBI's language for defining the grammar of an object language
together with inference rules is based on the logical framework LF;
pragmatically, we have adopted the concrete syntax of LF
specifications in Beluga, which is almost identical to Twelf's.  The
advantage is that specifications can be directly parsed and more
importantly type checked by Beluga, thereby eliminating many
syntactically correct but meaningless expressions.

Object languages are written according to the EBNF grammar in
Fig.~\ref{fig:gramlf}, which uses certain standard conventions: \verb!{a}!
means repeat a production zero or more times, and comments in the
grammar are enclosed between \verb!(*! and \verb!*)!.  The token
\verb!id!
refers to identifiers starting with a lower or upper case letter.
\begin{figure}[th]
\begin{verbatim}
sig      ::= {decl                      (* declaration *)
           | s_decl}                    (* schema declaration *)

decl     ::= id ":" tp "."              (* constant declaration *)
           | id ":" kind "."            (* type declaration *)

op_arrow ::= "->" | "<-"                (* A <- B same as B -> A *)

kind     ::= type
           | tp op_arrow kind           (* A -> K *)
           | "{" id ":" tp "}" kind     (* Pi x:A.K *)

tp       ::= id {term}                  (* a M1 ... M2 *)
           | tp op_arrow tp
           | "{" id ":" tp "}" tp       (* Pi x:A.B *)

term     ::= id                         (* constants, variables *)
           | "\" id "." term            (* lambda x. M *)
           | term term                  (* M N *)

s_decl    ::= schema s_id "=" alt_blk ";"

s_id      ::= id

alt_blk   ::= blk {"+" blk}                         

blk       ::= block id ":" tp {"," id ":" tp}  
\end{verbatim}
\caption{ORBI grammar for syntax, judgments, inference rules, and
  context schemas}
\label{fig:gramlf} 
\end{figure}
These grammar rules are basically the standard ones used both in Twelf
and Beluga and we do not
discuss them in detail here. We only note that while the presented
grammar permits general dependent types up to level $n$, ORBI
specifications will only use level $0$ and level $1$. Intuitively,
specifications at level $0$ define the syntax of a given object
language, while specifications at level $1$ (i.e., type families that are
indexed by terms of level $0$) describe the judgments and rules for a
given OL\@.  We exemplify the grammar relative to the example of
algorithmic vs.\ declarative equality.
For more example specifications, we refer the reader to our survey
paper \cite{companion} or to
\url{https://github.com/pientka/ORBI}.\footnote{The observant reader
  will have noticed that ORBI's concrete syntax for schemas differs
  from the one that we have presented in Sect.~\ref{sec:example}, in
so much that blocks are separated by commas and not by
semi-colons. This is forced on us by our choice to re-use Beluga's
parsing and checking tools.}

\paragraph{Syntax} 
An ORBI file starts in the \verb!Syntax! section with the declaration
of the constants used to encode the syntax of the OL in question, here
untyped lambda-terms, which are introduced with the declarations: 
\begin{verbatim}
%% Syntax
tm: type.
app: tm -> tm -> tm.
lam: (tm -> tm) -> tm.
\end{verbatim}

\noindent
The declaration introducing type \verb!tm! along with those of the constructors \verb!app! and
\verb!lam! fully specify the syntax of OL terms.  We represent binders
in the OL using binders in the HOAS meta-language. Hence the
constructor \verb!lam! takes in a function of type \verb!tm -> tm!.
 For example, the OL term $(\lam{x}{\lam{y}{\app{x}{y}}})$ is
represented as \verb!lam (\x. lam (\y. app x y))!, where
``$\verb-\-$'' is the binder of the metalanguage. Bound variables
found in the object language are not explicitly represented in the
meta-language.

\paragraph{Judgments and Rules} These are introduced as LF type
families (predicates)
in the \verb|Judgments| section followed by object-level inference
rules for these judgments in the \verb|Rules| section.  In our running example, we
have two judgments:
\begin{verbatim}
%% Judgments
aeq: tm -> tm -> type.
deq: tm -> tm -> type.
\end{verbatim}

\noindent Consider first the inference rule for
algorithmic equality for application, where the ORBI text is a
straightforward encoding of the rule:

\noindent
\begin{minipage}[c]{0.5\linewidth}
\begin{small}
\begin{verbatim}
ae_a: aeq M1 N1 -> aeq M2 N2 
   -> aeq (app M1 M2) (app N1 N2).
\end{verbatim}   
\end{small}
\end{minipage}
\begin{minipage}[c]{0.4\linewidth}
\[
\begin{array}{c}
~~~~~\infer[ae_a]{\Gamma \vdash \termeq{(\app{M_1}{M_2})}{(\app{N_1}{N_2})}}
{\Gamma \vdash \termeq{M_1}{N_1} & \Gamma \vdash \termeq{M_2}{N_2}}
\\\;
\end{array}
\]
\end{minipage}

\noindent
Uppercase letters such as \verb|M1| denote schematic variables,
which are implicitly quantified at the outermost level, namely
\verb|{M1:tm}|, as is commonly done for readability purposes in Twelf and
Beluga.

\noindent The binder case is  more interesting:

\noindent
\begin{minipage}[c]{0.6\linewidth}
\begin{small}
\begin{verbatim}
ae_l: ({x:tm} aeq x x -> aeq (M x) (N x)) 
   -> aeq (lam (\x. M x)) (lam (\x. N x)).
\end{verbatim}
\end{small}
\end{minipage}
\begin{minipage}[c]{0.1\linewidth}
\[
\begin{array}{c}
\infer[ae_l]{\Gamma \vdash \termeq{(\lam{x}{M})}{(\lam{x}{N})}}{
\Gamma, \istm{x}; \termeq{x}{x} \vdash \termeq{M}{N}} 
\\\;
\end{array}
\]
\end{minipage}

\noindent
We view the $\istm x$ assumption as the parametric assumption
\verb|x:tm|, while the hypothesis $\termeq x x$ (and its scoping) is
encoded within the embedded implication
\verb|aeq x x -> aeq (M x) (N x)| in the current (informal) signature
augmented with the dynamic declaration for \verb|x|. {As is
  well known,
  parametric assumptions and embedded implication are unified in the
  type-theoretic view.} Note that the ``variable'' case, namely rule
$ae_v$, is folded inside the binder case. We list here the rest of the
\verb|Rules| section:

\begin{verbatim}
%% Rules
de_a: deq M1 N1 -> deq M2 N2 -> deq (app M1 M2) (app N1 N2).
de_l: ({x:tm} deq x x -> deq (M x) (N x)) 
        -> deq (lam (\x. M x)) (lam (\x. N x)).
de_r: deq M M.
de_s: deq N M -> deq M N.
de_t: deq M L -> deq L N -> deq M N.
\end{verbatim}

\paragraph{Schemas} A {schema} declaration \verb!s_decl! is introduced
using the keyword \verb!schema!.  A \verb!blk! consists of one or more
declarations and \verb!alt_blk! describes \emph{alternating}
schemas. For example, schemas mentioned in Sect.~\ref{sec:example}
appear in the \verb!Schemas! section as:
\begin{verbatim}
%% Schemas
schema xG =  block (x:tm);
schema xaG = block (x:tm, u:aeq x x);
schema xdG = block (x:tm, u:deq x x);
schema daG = block (x:tm, u:deq x x, v:aeq x x);
\end{verbatim}  

To illustrate alternatives in contexts, consider extending our OL to
the polymorphically typed lambda-calculus, which includes a new type
\verb!tp! in the \verb!Syntax! section, and a new judgment:
\begin{verbatim}
atp: tp -> tp -> type.
\end{verbatim}
representing equality of types in the \verb!Judgments! section (as
well as type constructors and rules for well-formed types and type
equality, omitted here).  With this extension, the following two
examples replace the first two schemas in the \verb!Schemas! section.
\begin{verbatim}
schema xG  = block (x:tm) + block (a:tp);
schema xaG = block (x:tm, u:aeq x x) + block (a:tp, v:atp a a);
\end{verbatim}  

While we type-check the schema definitions using an extension of
the LF type checker (as implemented in Beluga), we
do not verify that the given schema definition is meaningful with
respect to the specification of the syntax and inference rules; in
other words, we do not perform ``world checking'' in Twelf lingo.

\paragraph{Definitions}
So far we have considered the specification language for encoding
formal systems. ORBI also supports declaring inductive definitions for
specifying context relations.  We start with the grammar
for inductive definitions (Fig.~\ref{fig:gramindef}).
Although we plan to provide syntax for specifying more general
inductive definitions, in this version of ORBI we \emph{only} define
\emph{context relations} inductively, that is $n$-ary predicates
between contexts of some given schemas. Hence the base predicate is of the
form \verb|id {ctx}| relating different contexts.
\begin{figure}[ht]
\begin{verbatim}
def_dec  ::= "inductive" id ":" r_kind "=" def_body ";"

r_kind   ::= "prop"
           | "{" id ":" s_id "}" r_kind

def_body ::= "|" id ":" def_prp {def_body}

def_prp  ::= id {ctx}
           | def_prp "->" def_prp

ctx      ::= [] | [id] | ctx "," id ":" blk
\end{verbatim}
\caption{ORBI grammar for inductive definitions describing context relations}
\label{fig:gramindef} 
\end{figure}
For example, the \verb!Definitions! section defines the relations
$\crel \ictx \actx$ and $\crel \actx \dctx$.  To illustrate, only the
former is shown below.
\begin{verbatim}
%% Definitions
inductive Rxa : {g:xG} {h:xaG} prop =
| Rxa_nl: Rxa [] []
| Rxa_cs: Rxa [g] [h] 
          -> Rxa [g, b:block (x:tm)] [h, b: block (x:tm, u:aeq x x)];
\end{verbatim}
This kind of relation can be translated fairly directly to inductive
n-ary predicates in systems supporting the proof-theoretic view. In
the type-theoretic framework underlying Beluga, inductive predicates
relating contexts correspond to recursive data types indexed by
contexts; in fact ORBI adopts Beluga's concrete syntax, so as to directly
type-check those definitions as well. Twelf's type theoretic
framework, however, is not rich enough to support inductive
definitions.

\subsection{Theorems and Directives in ORBI}
\label{ssec:logic}

While the elements of an ORBI specification detailed in the previous
subsection were relatively easy to define in a manner that is well
understood by all the different systems we are targeting, we
illustrate in this subsection those elements that are harder to
describe uniformly due to the different treatment and meaning of
contexts in the different systems.

\paragraph{Theorems}
We list the grammar for theorems in Fig.~\ref{fig:gramthm}.  Our
reasoning language includes a category \verb!prp! that specifies the logical
formulas we support. The base predicates include \verb!false,true!,
term equality, atomic predicates of the form \verb!id {ctx}!, which
are used to express context relations, and predicates of the form
\verb![ctx |- J]!, which represent judgments of an object language
within a given context.
Connectives and quantifiers include implication, conjunction,
disjunction, universal and existential quantification over terms, and
universal quantification over context variables.

\begin{figure}[ht]
\begin{verbatim}
thm      ::= "theorem" id ":" prp ";"

prp      ::= id {ctx}                     (* Context relation *)
           | "[" ctx  "|-" id {term} "]"  (* Judgment in a context *)
           | term "=" term                (* Term equality *)
           | false                        (* Falsehood *)
           | true                         (* Truth *)
           | prp "&" prp                  (* Conjunction *)
           | prp "||" prp                 (* Disjunction *)
           | prp "->" prp                 (* Implication *)
           | quantif prp                  (* Quantification *)

quantif  ::= "{" id ":" s_id "}"          (* universal over contexts *)
           | "{" id ":" tp "}"            (* universal over terms *)
           | "<" id ":" tp ">"            (* existential over terms *)
\end{verbatim}
\caption{ORBI grammar for theorems}
\label{fig:gramthm} 
\end{figure}

To illustrate, the reflexivity lemmas and completeness theorems for
both the G and R versions as they appear in the \verb!Theorems!
section are shown below. These theorems are a straightforward
encoding of those stated in Sect.~\ref{sec:example}.
\begin{verbatim}
%% Theorems
theorem reflG: {h:xaG}{M:tm} [h |- aeq M M];
theorem ceqG:  {g:daG}{M:tm}{N:tm} [g |- deq M N] -> [g |- aeq M N];

theorem reflR: {g:xG}{h:xaG}{M:tm}        Rxa [g] [h] -> [h |- aeq M M];
theorem ceqR:  {g:xdG}{h:xaG}{M:tm}{N:tm} Rda [g] [h] -> 
                                              [g |- deq M N] -> [h |- aeq M N];
\end{verbatim}

As mentioned, we do not type-check theorems; in particular, we do not
define the meaning of \verb![ctx |- J]!, since several interpretations
are possible.  In Beluga, every judgment \verb+J+ must be meaningful
within the given context \verb+ctx+; in particular, \emph{terms}
occurring in the judgment \verb+J+ must be meaningful in
\verb+ctx+. As a consequence, both parametric and hypothetical
assumptions relevant for establishing the proof of \verb+J+ must be
contained in \verb+ctx+. Instead of the local context view adopted in
Beluga, Twelf has one global ambient context containing all relevant
parametric and hypothetical assumptions. Systems based on proof-theory
such as Hybrid and Abella distinguish between assumptions denoting
eigenvariables (i.e., parametric assumptions), which live in a global
ambient context and proof assumptions (i.e., hypothetical assumptions),
which live in the context \verb+ctx+. While users of different systems
understand how to interpret \verb![ctx |- J]!, reconciling these
different perspectives in ORBI is beyond the scope of this paper. Thus
for the time being, we view theorem statements in ORBI as a kind of
\emph{comment}, where it is up to the user of a particular system to
determine how to translate them.

\paragraph{Directives}
\label{ssec:direct}

In ORBI, \emph{directives} are comments that help
the ORBI2X tools to generate target
representations of the ORBI specifications. The idea is reminiscent of
what  \emph{Ott}  \cite{ott} does to customize certain
declarations, e.g., the representation of variables, to the different
programming languages/proof assistants it supports.
The grammar for directives is listed in Fig.~\ref{fig:direct}.

\begin{figure}[th]
  \centering
\begin{verbatim}
dir       ::=  '%' sy_set what decl {"," decl} {dest} '.'
             | '%%' sepr '.'

sy_id     ::=  hy | ab | bel | tw 

sy_set    ::= '[' sy_id {',' sy_id} ']'

what      ::=  wf | explicit | implicit

dest      ::= 'in' ctx | 'in' s_id | 'in' id

sepr      ::=  Syntax | Judgments | Rules | Schemas | Definitions 
               | Directives | Theorems
\end{verbatim}  
  \caption{ORBI grammar for directives}
  \label{fig:direct}
\end{figure}

The \texttt{sepr} directives, such as \texttt{Syntax}, are simply
means to structure ORBI specifications.  Most of the other directives
that we consider in this version of ORBI are dedicated to help the
translations into proof-theoretical systems, although we also include
some to facilitate the translation of {theorems} to Beluga. The set of
directives is not intended to be complete and the meaning of
directives is system-specific.  The directives \verb|wf| and
\verb|explicit| are concerned with the asymmetry in the
proof-theoretic view between declarations that give typing
information, e.g., \verb|tm:type|, and those expressing judgments,
e.g., \verb|aeq:tm -> tm -> type|. In Abella and Hybrid, the former
may need to be reified in a judgment, in order to show that judgments
preserve the well-formedness of their constituents, as well as to
provide induction on the structure of terms; yet, in order to keep
proofs compact and modular, we want to minimize this reification and
only include them where necessary.  The \verb!Directives! section of
our sample specification includes, for example,
\begin{verbatim}
% [hy,ab] wf tm.
\end{verbatim}
which refers to the first line of the
\verb!Syntax! section where \verb!tm! is introduced, and indicates
that we need a predicate (e.g., \verb|is_tm|) to express
well-formedness of terms of type \verb|tm|.  Formulas expressing the
definition of this predicate are automatically generated from the
declarations of the constructors \verb|app| and \verb|lam|
with their types.

The keyword \verb|explicit| indicates when such well-formedness
predicates should be included in the translation of the declarations
in the \verb!Rules! section.  For example, the following
formulas both represent possible translations of the \verb|ae_l| rule
to proof-theoretic systems. We use Abella's concrete syntax to
exemplify:
\begin{verbatim}
aeq (lam M) (lam N) :- pi x\ is_tm x => aeq x x => aeq (M x) (N x).
aeq (lam M) (lam N) :- pi x\ aeq x x => aeq (M x) (N x).
\end{verbatim}
where the typing information is explicit in the first and implicit in
the second.  By default, we choose the latter, that is well-formed
judgments are assumed to be \emph{implicit}, and require a directive
if the former is desired. 
Consider, for example, that we want to conclude that whenever a
judgment is provable, the terms in it are well-formed, e.g., if
\verb|aeq M N| is provable, then so are \verb|is_tm M| and
\verb|is_tm N|.  Such a lemma is indeed provable in Abella and Hybrid
from the \emph{implicit} translation of the rules for \verb|aeq|.
Proving a similar lemma for the \verb|deq| judgment, on the other
hand, requires some strategically placed explicit well-formedness
information.  In particular, the two directives:
\begin{verbatim}
% [hy,ab] explicit (x : tm) in de_l.
% [hy,ab] explicit (M : tm) in de_r.
\end{verbatim}
require the clauses \verb|de_l| and \verb|de_r| to be translated to the
following formulas:
\begin{verbatim}
deq (lam M) (lam N) :- pi x\ is_tm x => deq x x => deq (M x) (N x).
deq M M :- is_tm M. 
\end{verbatim}

The case for schemas is analogous.  In the systems based on
proof-theoretic approaches, contexts are typically represented using
lists and schemas are translated to
unary inductive predicates that verify that these lists have a
particular regular structure. We again leave typing information
implicit in the translation unless a directive is included.  For
example, the \verb|xaG| schema with no associated directive will be
translated to the following inductive definition in Abella:
 \begin{verbatim}
Define aG : olist -> prop by
  xaG nil;
  nabla x, xaG (aeq x x :: As) := xaG As.
\end{verbatim}
The directive \verb|% [hy,ab] explicit (x : tm) in daG|
will yield this Hybrid definition:
\begin{verbatim}
Inductive daG : list atm -> Prop :=
| nil_da : daG nil
| cns_da : forall (Gamma:list atm) (x:uexp),
    proper x -> daG Gamma -> daG (is_tm x :: deq x x :: aeq x x :: Gamma).
\end{verbatim}

Similarly, directives in context relations, such as:
\begin{verbatim}
% [hy,ab] explicit (x : tm) in g in Rxa.
\end{verbatim}
also state which well-formedness annotations to make explicit in the
translated version. In this case, when translating the definition of
\verb!Rxa! in the \verb!Definitions! section, they are to be kept in
\verb|g|, but skipped in \verb|h|.

Keeping in mind that we consider the notion of directive
\emph{open} to cover other benchmarks and different systems, we offer
some speculation about directives that we may need to translate
theorems for the examples and systems that we are considering.
For example, theorem \verb|reflG| is proven by induction over
\verb+M+. As a consequence, \verb+M+ must be explicit.
\begin{verbatim}
% [hy,ab,bel] explicit (M : tm) in h in reflG.
\end{verbatim}
\noindent
The ORBI2Hybrid and ORBI2Abella tools will interpret the directive by
adding an explicit assumption,
as illustrated by the result of the ORBI2Abella translation:
\begin{verbatim}
forall H M, xaG H -> {H |- is_tm M} -> {H |- aeq M M}.
\end{verbatim}
In Beluga, the directive is interpreted as:
\begin{verbatim}
{h:xaG} {M:[h |- tm]} [h |- aeq M M].
\end{verbatim}
where \verb!M! will have type \verb!tm! in the context
\verb!h!. Moreover, since the term \verb!M! is used in the
judgment \verb!aeq! within the context \verb!h!, we
associate \verb!M!  with an identity substitution, which is not displayed.  In short, the directive allows us to lift the type
specified in ORBI to a contextual type that is meaningful in Beluga.
In fact, Beluga always needs additional information on how to
interpret terms---are they closed or can they depend on a given
context? For translating \verb+symG+ for example, we use the following
directive to indicate the dependence on the context:
\begin{verbatim}
% [bel] implicit (M : tm), (N : tm) in h in symG.
\end{verbatim}

\subsection{Guidelines}
\label{ssec:guidelines}

In addition, we introduce a set of \emph{guidelines} for ORBI specification
writers, with the goal of helping translators generate output that is
more likely to be accepted by a specific system.  ORBI 0.1 includes
four such guidelines, which are motivated by the desire to avoid putting too
many constraints in the grammar rules.  First, as we have seen in our
examples, we use as a convention that free
variables which denote schematic variables in rules are written using upper
case identifiers; we use lower case identifiers for eigenvariables in
rules and for context variables. Second, while the grammar does not restrict what types we can quantify
over, the intention is that we quantify over types of level-0, i.e., objects of
the syntax level, only. Third, in order to more easily accommodate
systems without dependent types, \verb!Pi! should not be used when
writing non-dependent types; an arrow should be used instead.  (In
LF, for example, \verb!A -> B! is an abbreviation for \verb!Pi x:A.B!
for the case when \verb!x! does not occur in \verb!B!.  Following this
guideline means favoring this abbreviation whenever it applies.)
Fourth, when writing a context (grammar
\verb|ctx|), distinct variable names should be used in different
blocks.

\section{Related Work}
\label{sec:related}

Our approach to structuring contexts of assumptions takes its
inspiration from Martin-L\"of's theory of judgments, especially in the way it has been realized in 
Edinburgh LF\@. However, our formulation owes more to Beluga's
type theory, where contexts are first-class citizens, than to the
notion of \emph{regular world} in Twelf.

The creation and sharing of a library of benchmarks has proven to be
very beneficial to the field it represents. The brightest example is
\emph{TPTP} \cite{TPTP}, whose influence on the development, testing and
evaluation of automated theorem provers cannot be
underestimated. Clearly our ambitions are much more limited. We have
also taken some inspiration from its higher-order extension \emph{THF0}
\cite{THF0}, in particular in its construction in stages. 

The success of {TPTP} has spurred other benchmark suites in related
subjects, see for example \emph{SATLIB} \cite{SATLIB}; however, the only one
concerned with induction is the \emph{Induction Challenge Problems}
(\url{http://www.cs.nott.ac.uk/~lad/research/challenges}), a
collection of examples geared to the \emph{automation} of inductive
proof. The benchmarks are taken from arithmetic, puzzles, functional
programming specifications, etc.\ and as such have little connection
with our endeavor.
On the other hand, the examples mentioned earlier coming from Twelf's
wiki, Abella's library, Beluga's distribution, and Hybrid's web page
contain a set of examples that highlight the issues around binders. As such
they are prime candidates to be included in ORBI.

Other projects have put forward LF as a common ground:
the goal of \emph{Logosphere}'s (\url{http://www.logosphere.org}) was the
design of a representation language for logical formalisms, individual
theories, and proofs, with an interface to other theorem proving
systems that were somewhat connected, but the project never
materialized. \emph{SASyLF} \cite{SASyLF} originated as a tool to
teach programming language theory: the user specifies the syntax,
judgments, theorems \emph{and} proofs thereof (albeit limited to
\emph{closed} objects) in a paper-and-pencil HOAS-friendly way and the
system converts them to totality-checked Twelf code. The capability to
express and share proofs is of obvious interest to us, although such
proofs, being a literal proof verbalization of the corresponding Twelf type
family, are irremediably verbose. Finally, work on modularity in LF
specifications \cite{RabeS09} is of critical interest to give more
structure to ORBI files.

\emph{Why3} (\url{http://why3.lri.fr}) is a software verification
platform that intends to provide a front-end to third-party theorem
provers, from proof assistants such as Coq to SMT-solvers.  To this
end Why3 provides a first-order logic with rank-1 polymorphism,
recursive definitions, algebraic data types and inductive predicates
\cite{why3}, whose specifications are then translated to the several
systems that Why3 supports. Typically, those translations are
forgetful, but sometimes, e.g., with respect to Coq, they add some
annotations, for example to ensure non-emptiness of types. Although we
are really not in the same business as Why3, there are several ideas
that are relevant; to name one, the notion of a \emph{driver}, that
is, a configuration file to drive transformations specific to a
system. Moreover, Why3 provides an API for users to write and
implement their own drivers and transformations.

\emph{Ott} \cite{ott} is a highly engineered tool for ``working
semanticists,'' allowing them to write programming language
definitions in a style very close to paper-and-pen specifications; 
then  those are compiled into \LaTeX\ and, more interestingly, into
proof assistant code, currently supporting Coq, Isabelle/HOL, and
HOL\@. Ott's metalanguage is endowed with a rich theory of binders,
but at the moment it favors the ``concrete'' (non $\alpha$-quotiented)
representation, while providing support for the nameless
representation for a single binder. Conceptually, it would be natural
to extend Ott to generate ORBI code, as a bridge for Ott to support
HOAS-based systems. Conversely, an ORBI user would benefit from having
Ott as a front-end, since the latter view of grammar and judgment
seems at first sight general enough to support the notion of schema
and context relation.

In the category of environments for programming language descriptions,
we mention \emph{PLT-Redex}
\cite{PLTbook} and also the \emph{K} framework \cite{RosuK}. In both, several
large-scale language descriptions have been specified and
tested. However, none of those systems has any support for binders, let alone
context specifications, nor can any meta-theory be formally verified.

Finally, there is a whole research area dedicated to the handling and
sharing of mathematical content (\emph{MMK}
\url{http://www.mkm-ig.org}) and its representation (\emph{OMDoc}
\url{https://trac.omdoc.org/OMDoc}), which is only very loosely
connected to our project.

\section{Conclusion}
\label{sec:concl}

We have presented the preliminary design of a language, and more
generally, of a common infrastructure for representing challenge
problems for HOAS-based logical frameworks.  The common notation
allows us to express the syntax of object languages that we wish to
reason about, as well as the context schemas, the judgments and
inference rules, and the statements of benchmark theorems.

We strongly believe that the field has matured enough to benefit from
the availability of a set of benchmarks on which qualitative and
hopefully quantitative comparison can be carried out.  We hope that
ORBI will foster sharing of examples in the community and provide a
common set of examples.  We also see our benchmark repository as a
place to collect and propose ``open'' challenge problems to push the
development of meta-reasoning systems.

The challenge problems also play a role in allowing us, as designers
and developers of logical frameworks, to highlight and explain how the
design decisions for each individual system lead to differences in
using them in practice. Additionally, our benchmarks aim to provide a
better understanding of what practitioners should be looking for, as
well as help them foresee what kind of problems can be solved
elegantly and easily in a given system, and more importantly, why this
is the case. Therefore the challenge problems provide guidance for
users and developers in better comprehending differences and
limitations. Finally, they serve as an excellent regression suite.

The description of ORBI presented here is best thought of as a
stepping stone towards a more comprehensive specification language,
much as \emph{THF0} \cite{THF0} has been extended to the more
expressive formalism $THF_i$, adding for instance, rank-1
polymorphism. Many are the features that we plan to provide in the
near future, starting from general (monotone) \emph{(co)inductive}
definitions; currently we only relate contexts, while it is clearly
desirable to relate arbitrary well-typed terms, as shown for example
in \cite{Cave:POPL12} and \cite{GacekMillerNadathur:JAR12} with
respect to normalization proofs. Further, it is only natural to
support infinite objects and behavior.  However, full support for
(co)induction is a complex matter, as it essentially entails fully
understanding the relationship between the proof-theory behind Abella
and Hybrid and the type theory of Beluga. Once this is in place, we
can ``rescue'' ORBI theorems from their current status as comments and
even include proof sketches in ORBI.

Clearly, there is a significant amount of implementation work ahead,
mainly on the ORBI2X tools side, but also on the practicalities
of the benchmark suite. Finally, we would like to open up the
repository to other styles of formalization such as nominal, locally
nameless, etc.

\end{document}